\documentclass[aps,prb,twocolumn,groupedaddress,showpacs,floatfix]{revtex4}

\usepackage{graphicx}
\usepackage{subfigure}
\usepackage{amsmath}
\usepackage{amssymb}
\usepackage{bm}
\usepackage{dcolumn}
\usepackage{nicefrac}

\newcommand{\zn}{ZnCu$_3$(OH)$_6$Cl$_2$}
\newcommand{\ka}{kagom{\'e}}

\begin{document}

\title{First-principles determination of Heisenberg Hamiltonian parameters for the spin-1/2 kagom{\'e} antiferromagnet ZnCu$_3$(OH)$_6$Cl$_2$}

\author{Harald O. Jeschke}
\email{jeschke@itp.uni-frankfurt.de}
\affiliation{Institut f\"ur Theoretische Physik, Goethe-Universit\"at Frankfurt, Max-von-Laue-Strasse 1, 60438 Frankfurt am Main, Germany}

\author{Francesc Salvat-Pujol}
\affiliation{Institut f\"ur Theoretische Physik, Goethe-Universit\"at Frankfurt, Max-von-Laue-Strasse 1, 60438 Frankfurt am Main, Germany}

\author{Roser Valent\'\i}
\affiliation{Institut f\"ur Theoretische Physik, Goethe-Universit\"at Frankfurt, Max-von-Laue-Strasse 1, 60438 Frankfurt am Main, Germany}

\date{\today}

\begin{abstract}
  Herbertsmithite ({\zn}) is often discussed as the best realization
  of the highly frustrated antiferromagnetic {\ka} lattice known
  so far. We employ density functional theory calculations to
  determine eight exchange coupling constants of the underlying
  Heisenberg Hamiltonian. We find the nearest neighbour coupling $J_1$
  to exceed all other couplings by far. However, next-nearest
  neighbour {\ka} layer couplings of $0.019J_1$ and interlayer
  couplings of up to $-0.035J_1$ slightly modify the perfect
  antiferromagnetic {\ka} Hamiltonian. Interestingly, the largest
  interlayer coupling is ferromagnetic even without Cu impurities in
  the Zn layer. In addition, we validate our DFT approach by applying
  it to kapellasite, a polymorph of herbertsmithite which is known
  experimentally to exhibit competing exchange interactions.
\end{abstract}


\maketitle


Quantum spin liquids have fascinated physicists for decades as this
exotic ground state constitutes a novel state of
matter~\cite{Balents2010}. The magnetic moments in a spin liquid do
not order even at extremely low temperature due to a high degree of
frustration in the magnetic system. Typical examples for lattices that
lead to frustration of antiferromagnetic interactions are triangular,
pyrochlore and {\ka} lattices. While experimental realizations of
quantum spin liquids have long been scarce, in particular the
discovery~\cite{Shores2005} of the perfect {\ka} lattice realization
in herbertsmithite has led to considerable
excitement~\cite{Lee2008}. In the eight years since the discovery of
the $S=\nicefrac{1}{2}$ {\ka} antiferromagnet nature of {\zn},
numerous experiments have been performed to ascertain the spin liquid
ground state of herbertsmithite, its properties and
excitations~\cite{Mendels2010,Mendels2011}. In particular,
measurements of the magnetic susceptibility~\cite{Helton2007} show
antiferromagnetic couplings of the order $J\approx 17$~meV ($\sim 190$
K) and no magnetic ordering down to 50~mK. Muon spin rotation
measurements~\cite{Mendels2007} confirm the absence of magnetic
ordering and inelastic neutron scattering
experiments~\cite{Vries2009,Han2012} find that fractionalized quantum
excitations are present in {\zn}. More recently, non-ideality of the
realization of the {\ka} Heisenberg Hamiltonian in herbertsmithite due
to additional interactions and in the form of site disorder has been
the focus of many studies~\cite{Mendels2011}.
While defects within the {\ka} layer are detected in nuclear magnetic
resonance~\cite{Olariu2008} but not in recent x-ray scattering
measurements~\cite{Freedman2010}, Cu impurities on interlayer Zn sites
seem to play a role~\cite{Freedman2010}.  Low temperature deviations
between theory for the {\ka} antiferromagnet and experimental
susceptibilities as well as anisotropies in thermodynamic
quantities~\cite{Han2012b} point to a small nonzero
Dzyaloshinskii-Moriya
interaction~\cite{Rigol2007,Messio2010}. Evidence of this interaction
has been found in electron spin resonance
measurements~\cite{Zorko2008,Shawish2010}.  However, the experimental
and theoretical discussion about the Hamiltonian correctly describing
herbertsmithite is far from settled.

\begin{figure}[hbt]
\includegraphics[width=0.45\textwidth]{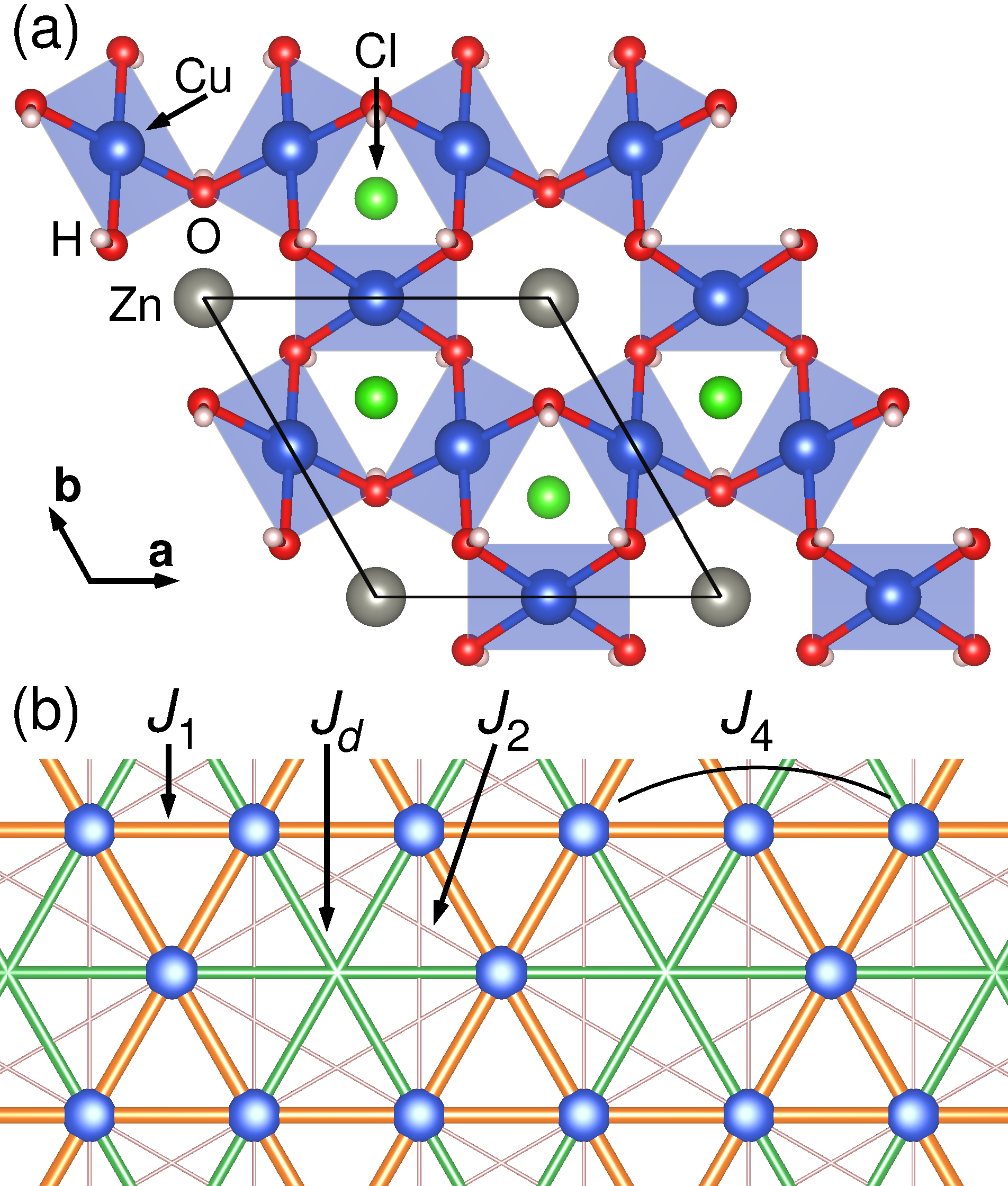}
\caption{(Color online) (a) Crystal structure of kapellasite, viewed
  along the {\bf c} direction. (b) Kagom{\'e} lattice formed by the Cu
  sites in (a). Note that $J_d$ and $J_4$ exchange paths both
  correspond to a distance of $6.3$~{\AA} but while $J_4$ points
  precisely along a nearest neighbour bond, $J_d$ cuts diagonally
  across a Cu hexagon with nonmagnetic Zn in the center.
}\label{fig:kapellasite}
\end{figure}

\begin{table}[htb]
  \caption{ Exchange coupling constants for {\zn} (kapellasite) determined from total energies of
    five different spin configurations in a $2\times2\times1$ supercell.
  }\label{tab:Jkapellasite221U6}
\begin{ruledtabular}
\begin{tabular}{cD{.}{.}{1.5}cD{.}{.}{1.1}}
name& \multicolumn{1}{c}{$d_{Cu-Cu}$}&type&\multicolumn{1}{c}{$J_i$~(K)}\\
&&&\multicolumn{1}{c}{$U\!=\!6\,$eV}\\\hline
$J_1$&3.15& {\ka} nn& -14.2\\
$J_2$&5.45596&{\ka} 2nd nn&-0.7\\
$J_4$&6.3 & {\ka} 3rd nn& -0.3\\
$J_d$&6.3 & {\ka} 3rd nn& 24.0
\end{tabular}
\end{ruledtabular}
\end{table}

\begin{figure*}[hbt]
\includegraphics[width=0.8\textwidth]{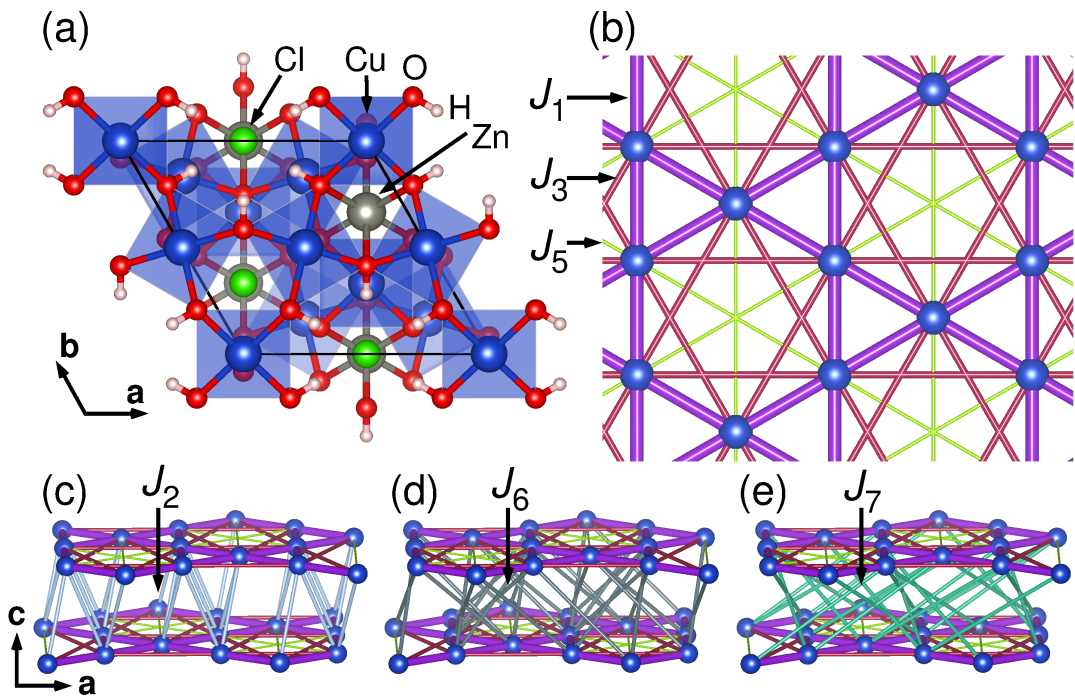}
\caption{(Color online) (a) Crystal structure of herbertsmithite
  {\zn}, viewed along the {\bf c} direction. (b) Kagom{\'e} lattice
  formed by the Cu sites in (a). Exchange paths between nearest, next
  nearest and third nearest neighbours within the {\ka} lattice are
  shown. (c)-(e) show three inter-{\ka} layer exchange
  pathways. }\label{fig:structure}
\end{figure*}

\begin{figure}[hbt]
\includegraphics[width=0.5\textwidth]{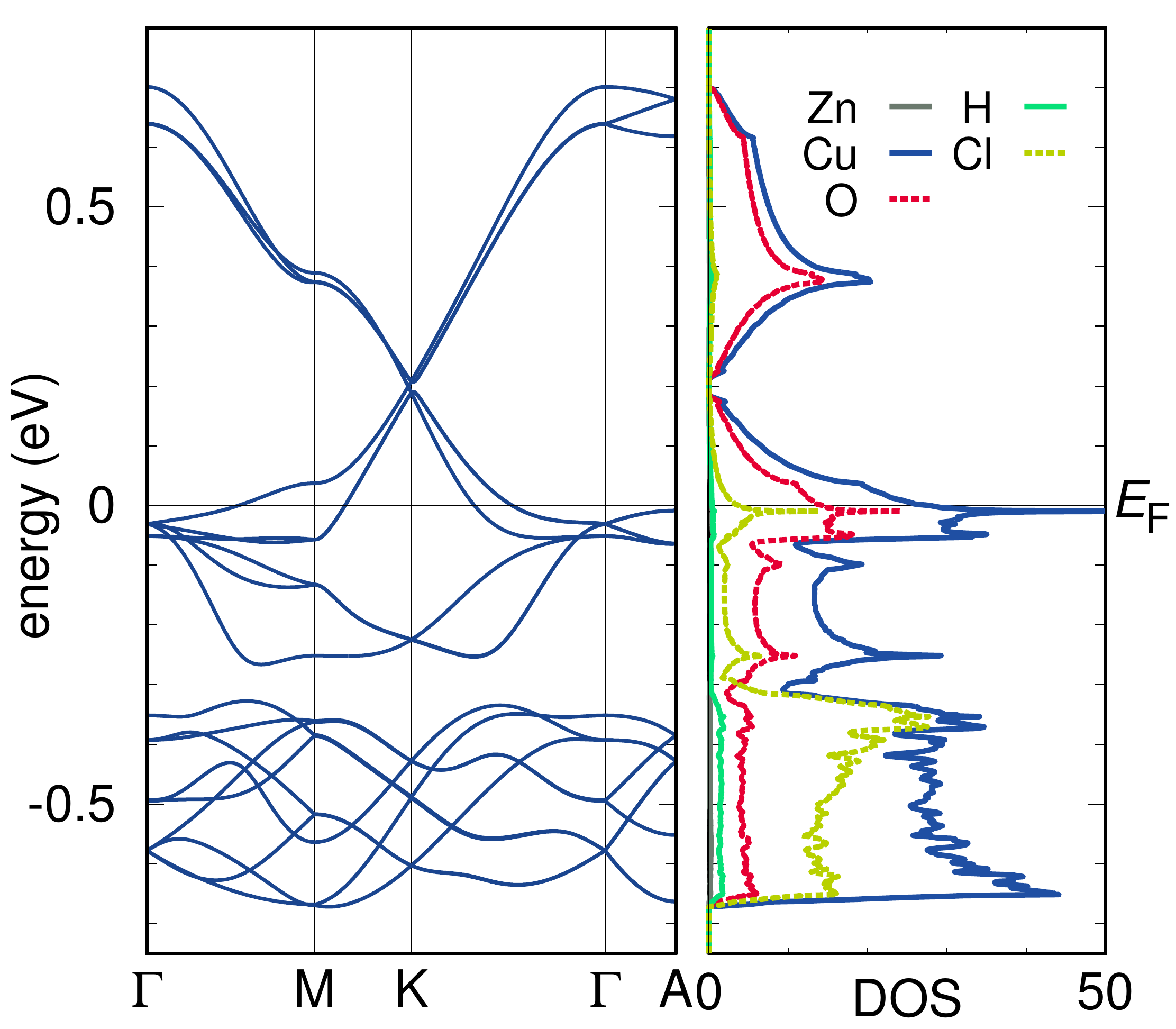}
\caption{(Color online) Band structure and density of states of {\zn}
  calculated with GGA exchange correlation functional. High symmetry
  points of the $P\,\bar{3}m$ space group are
  $M=(\nicefrac{1}{2},0,0)$, $K=(\nicefrac{1}{3},\nicefrac{1}{3},0)$
  and $A=(0,0,\nicefrac{1}{2})$ in units of the reciprocal lattice
  vectors. DOS is given in states per eV per unit cell (containing 3
  formula units).}\label{fig:bsdos}
\end{figure}

\begin{table}[htb]
  \caption{ 
    Exchange coupling constants for {\zn} (herbertsmithite) determined 
    from total energies of nine different spin configurations. Energies 
    were calculated with GGA+U functional at $U=6$~eV, $J=1$~eV and 
    with atomic limit double counting correction.}\label{tab:J}
\begin{ruledtabular}
\begin{tabular}{cD{.}{.}{1.5}cD{.}{.}{1.1}}
name& \multicolumn{1}{c}{$d_{Cu-Cu}$}&type&\multicolumn{1}{c}{$J_i$~(K)}\\
&&&\multicolumn{1}{c}{$U\!=\!6\,$eV}\\\hline
\multicolumn{4}{c}{{\bf {\ka} layer couplings}}\\\hline
$J_1$&3.4171& {\ka} nn&182.4  \\
$J_3$&5.91859&{\ka} 2nd nn&  3.4\\
$J_5$&6.8342 & {\ka} 3rd nn& -0.4\\\hline
\multicolumn{4}{c}{{\bf interlayer couplings}}\\\hline
$J_2$&5.07638& interlayer 1st nn&  5.3\\
$J_4$&6.11933& interlayer 2nd nn& -1.5\\
$J_6$&7.00876& interlayer 3rd nn&  -6.4\\
$J_7$&8.51328& interlayer 4th nn&  3.0\\
$J_9$&9.17347& interlayer 6th nn&  2.5\\
\end{tabular}
\end{ruledtabular}
\end{table}

Therefore, we undertake an effort to determine the parameters of the
underlying Heisenberg Hamiltonian using all-electron density
functional theory methods. We will show in this Letter that the
exchange coupling constants from first principles corroborate that
{\zn} is a near perfect realization of a {\ka} antiferromagnet with a
dominant coupling of $J_1=182$~K. However, there are small corrections
to this picture: A next-nearest neighbour coupling in the {\ka} layer
of $0.019J_1$ and in particular some interplanar couplings between
$-0.035J_1$ and $0.029J_1$ could actually be relevant for the nature
and excitations of the spin liquid ground state in herbertsmithite.

We perform density functional theory calculations with the full
potential local orbital (FPLO) basis set~\cite{Koepernik1999} using
generalized gradient approximation (GGA)~\cite{Perdew1996} and GGA+U
functionals. The exchange couplings, $J_i$, are obtained from total
energy calculations for different Cu spin configurations in supercells
of various sizes~\cite{Jeschke2011}.  Before proceeding to
herbertsmithite, we test our methods on kapellasite, a polymorph of
herbertsmithite, which has been investigated before, both
theoretically~\cite{Janson2008} and experimentally~\cite{Fak2012}. We
use the structure of kapellasite as given in
Ref.~\onlinecite{Krause2006} and determine the hydrogen position by
relaxation~\cite{hydrogen}. The structure is shown in
Figure~\ref{fig:kapellasite}. We create two different supercells: A
$2\times1\times2$ supercell with $P\,\bar{1}$ symmetry and 10
inequivalent Cu positions with the purpose of resolving four
interlayer couplings, and a $2\times2\times1$ supercell with $P\,1$
symmetry that provides symmetry inequivalent 3rd nearest neighbours in
the {\ka} plane. Using GGA+U with $U=6$~eV and $J=1$~eV, we find two
significant couplings, $J_1=-14.2$~K and $J_d=24.0$~K (see
Table~\ref{tab:Jkapellasite221U6}). Other couplings like $J_2$ are
significantly smaller (around 1~K), and we find interlayer couplings
to be negligible (see Appendix~\ref{appA}).  Note that the numbers we give are
converged to sub-Kelvin precision with our choice of spin
configurations; however, different sets of spin configurations will
lead to slightly different values so that we estimate the uncertainty
of the exchange constants for both compounds discussed in this work to
be around 1~K.  Systematic studies on the influence of the choice of
exchange and correlation functional and other technical variations in
the DFT determination of exchange couplings have been performed by
some of us in
Refs.~\onlinecite{Jeschke2011,Foyevtsova2011}. Table~\ref{tab:Jkapellasite221U6}
is in very good agreement with the observation of
Ref.~\onlinecite{Fak2012} that experimental data are compatible with a
$J_1$-$J_d$ model with $J_1=-15.0(4)$~K and $J_d=12.7(3)$~K. Note that
the ratio between $J_1$ and $J_d$ depends on the choice of $U$ as the antiferromagnetic $J_d$  in
particular is inversely proportional to
$U$ (see Appendix~\ref{appA}).  Even more recently, the high temperature
series expansion method was refined to fit both magnetic
susceptibility and specific heat data, yielding the set of parameters
$J_1=-12$~K, $J_2=-4$~K, $J_d=15.6$~K~\cite{Bernu2013}.  Thus, we can
proceed with some confidence to analyze the Heisenberg Hamiltonian
parameters of herbertsmithite.

We use the structure of {\zn} (herbertsmithite) with $R\,\bar{3}m$
space group determined by Shores {\it et al.}~\cite{Shores2005} which
is shown in Figure~\ref{fig:structure}~(a). A big difference with
respect to the polymorph kapellasite is that Zn is now between {\ka}
layers rather than in the centers of its hexagons. In
Figure~\ref{fig:bsdos}, we present the bandstructure and density of
states. At the Fermi level, we find Cu $3d$ states which hybridize
with O $2p$ and Cl $3p$ states. As expected, Zn plays no role at
$E_F$. Note that the Dirac-point-like feature at $K$ for an energy of
0.2~eV above the Fermi level becomes an avoided crossing with a tiny
gap in a fully relativistic calculation.  Based on our experience with
azurite, another complex quantum spin system containing Cu$^{2+}$
ions~\cite{Jeschke2011} and the fact that kapellasite, the polymorph
of herbertsmithite briefly analyzed above, was shown to have longer
ranged competing interactions~\cite{Fak2012}, we determine all
exchange constants up to Cu-Cu distances of 8.6~{\AA}. In order to
allow for determination of the diagonal coupling in the {\ka} lattice,
$J_5$, we double the unit cell along {\bf a} and prepare a structure
with $P\,m$ space group and 12 inequivalent Cu sites. As appropriate
for Cu, we employ a GGA+U exchange correlation functional with
$U=6$~eV, $J=1$~eV and atomic-limit double-counting
correction~\cite{Jeschke2011}. Total energies for nine different spin
configurations allow us to calculate the eight exchange coupling
constants listed in Table~\ref{tab:J}. The three couplings within the
{\ka} layer are shown in Figure~\ref{fig:structure}~(b), and the
geometry of the three most important interlayer couplings is presented
in Figure~\ref{fig:structure}~(c)-(e). To avoid confusion, we number
the coupling constants $J_i$ strictly according to ascending Cu-Cu
distances.  While the absolute values of the exchange constants
obtained from these calculations are dependent on the choice of the
$U$ value in the GGA+U calculations, as already pointed out for
kapellasite, one expects that the antiferromagnetic exchange constants
follow a $1/U$ law while the ferromagnetic exchange constants should
be less sensitive to the $U$ value. This trend is also observed in the
case of herbertsmithite when we compare the exchange constants
obtained for $U=6$~eV, $U=7$~eV and $U=8$~eV (see Appendix~\ref{appB}).  In
order to have also a quantitative description of the exchange
constants and not only ratios, we take as reference the
results obtained for $U=6$~eV, guided by the experience with
other copper-based materials as mentioned above.

Within the {\ka} layer, the most important correction to the presently
discussed Hamiltonian for herbertsmithite is the
next-nearest-neighbour coupling $J_3$. Theoretical investigations of
the {\ka} lattice with nearest and next-nearest neighbour interactions
indicate that the nature of the spin liquid ground state could depend
on such a next nearest coupling~\cite{Tay2012,Iqbal2012}. Messio {\it
  et al.}~\cite{Messio2012} have even extended the range of the
couplings to the 3rd nearest neighbour across a hexagon; our set of
parameters would put herbertsmithite in the $q=0$ spin liquid phase,
in agreement with Ref.~\onlinecite{Han2012}. Very recently, the
Heisenberg model on the {\ka} lattice with nearest and next-nearest
neighbour couplings has been studied with a pseudofermion functional
renormalization group method~\cite{Suttner2013}; very good agreement
with the inelastic neutron scattering experiment of
Ref.~\onlinecite{Han2012} is reached for the next nearest neighbour
interaction in the plane $J_3=0.017J_1$ which is very close to our
value.

Now we come to the interlayer couplings. First of all, it is important
to note that while each Cu site has four interactions via $J_1$, four
via $J_3$ and six via $J_5$, the interlayer bonds are numerous: There
are four $J_2$ bonds, six $J_4$ bonds, eight $J_6$ bonds, six $J_7$
bonds and six $J_9$ bonds. Interestingly, we find $J_2$ to be an
antiferromagnetic interlayer coupling of size $0.029J_1$, and $J_6$ a
ferromagnetic interlayer coupling of size $-0.035J_1$.  Previous
studies based on the spin-rotation-invariant Green's function method
showed that a stacked kagom\'e system remains short-range ordered
independent of the sign and strength of the interlayer
coupling~\cite{Schmalfuss2004}.  We have performed a first test of the
relevance of the interlayer couplings for susceptibility and specific
heat using high temperature series expansion~\cite{Schmidt2011}. This
method has been very useful to discuss the {\ka} lattice Heisenberg
model~\cite{Misguich2007} as well as various additional
terms~\cite{Singh2009}.  We find that at least in the region of
applicability of this method the effect of interlayer couplings is
noticeable.  We hope that our results inspire more precise manybody
calculations that could establish the consequences of interlayer
couplings for the low temperature properties of herbertsmithite.

In summary, our {\it ab initio}-based analysis of the Cu-Cu exchange
coupling constants in kapellasite and herbertsmithite provides a
detailed description of these materials. Our results for the dominant
interactions are in excellent agreement with experiments.  Moreover,
we are able to resolve the strength and sign of weaker, but not
negligible, exchange interactions that were not known up to now and
are important for understanding the behavior of these materials at low
temperatures.  Both polymorphs, even though they are realizations of a
perfect {\ka} lattice, show a few remarkable differences. The nearest
neighbor Cu-Cu exchange interaction is strongly antiferromagnetic in
herbertsmithite ($\sim 190$ K) and weakly ferromagnetic in kapellasite
($\sim -13$ K) due to the fact that the Cu-O-Cu angle in
herbertsmithite is $119^\circ$ compared to $106^\circ$ in kapellasite.
Kapellasite shows a significant antiferromagnetic 4th nearest neighbor
coupling along the diagonal of the Cu hexagon ($J_d$) which is
negligible in herbertsmithite since the exchange path in kapellasite
is through the in-plane Zn situated in the center of the
hexagons. Also, the stacking of the {\ka} layers in both polymorphs is
crucial for understanding the interlayer exchange couplings. In
kapellasite, the {\ka} layers are stacked in a similar fashion as in
the layered TiOCl~\cite{Zhang2008} or
Cs$_2$CuCl$_4$~\cite{Foyevtsova2011} where interactions are mostly of
van der Waals nature. In this situation, the interlayer couplings are
comparatively small (see Appendix~\ref{appB}). In contrast, in herbertsmithite
the interlayer Cu-Cu couplings are partly through Zn orbitals.  This
leads to relatively significant antiferromagnetic ($J_2$) and
ferromagnetic ($J_6$) interlayer couplings. Nevertheless, the ratio
between the dominant intralayer coupling $J_1$ and the dominant
interlayer coupling remains large enough for this system to be
considered a very good realization of a two-dimensional {\ka} lattice
and only at low temperatures should the smaller $J_i$ become
important.  This and the importance of couplings other than the
dominant ones for the spin-liquid behavior in these materials should
be investigated in the future. In particular, it would be interesting
to determine also the couplings of the Dzyaloshinskii-Moriya and ring exchange terms in
the Hamiltonian from first principles.

This work was supported in part by the National Science Foundation
under Grant No. NSF PHY11-25915, by the DFG through TRR/SFB 49, by the
Beilstein Institut through NanoBiC and by the Helmholtz Association
through HA216/EMMI. We would like to thank A. Honecker, C. Lhuillier,
P. Mendels and R. Moessner for useful discussions. Structure figures
were prepared with VESTA$\,$3~\cite{Momma2011}.

\appendix
\section{Details for exchange constants of kapellasite}\label{appA}

\begin{figure}[hbt]
\includegraphics[width=0.45\textwidth]{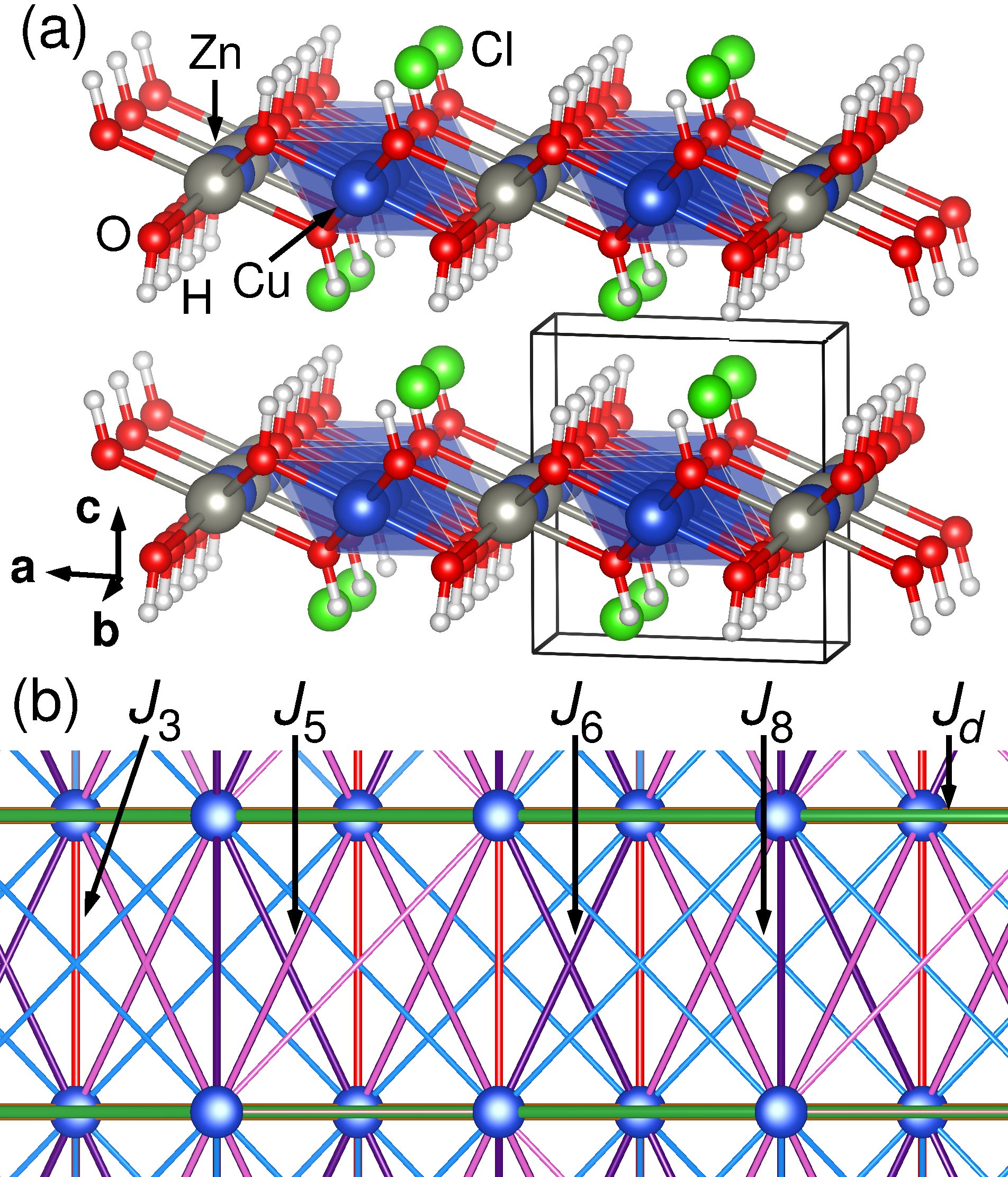}
\caption{(Color online) (a) Side view of the crystal structure of kapellasite (approximately
  along {\bf b} direction). (b) Interlayer exchange paths for the Cu
  sites in (a). 
}\label{fig:kapellasiteside}
\end{figure}

\begin{figure}[hbt]
\includegraphics[width=0.45\textwidth]{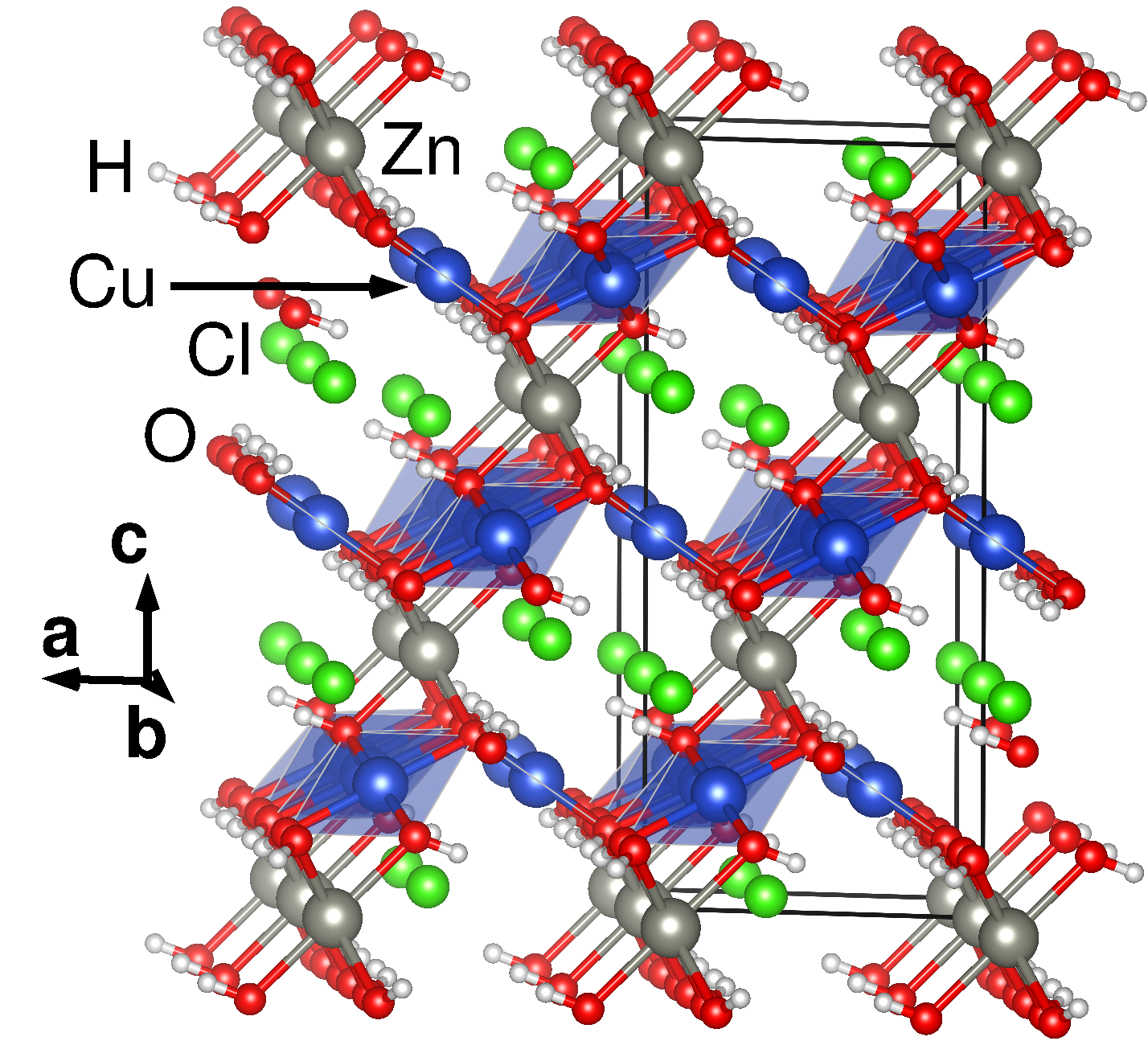}
\caption{(Color online) Side view of the crystal structure of herbertsmithite (approximately
  along {\bf b} direction). 
}\label{fig:herbertsmithiteside}
\end{figure}

In Tables~\ref{tab:Jkapellasite221} and \ref{tab:Jkapellasite212}, we
provide the results of total energy calculations with GGA+U functional
using different values of $U$. The $2\times1\times2$ supercell used in
the calculation for Table~\ref{tab:Jkapellasite212} allows resolution
of four interlayer couplings of kapellasite. They are all very small
which is not surprising considering the van der Waals gap between the
layers of kapellasite (see Figure~\ref{fig:kapellasiteside}). This is a
significant difference to the polymorph herbertsmithite that has {\ka}
layers coupled in the third dimension via O-Zn-O bonds (see
Figure~\ref{fig:herbertsmithiteside}).

\section{Details for exchange constants of herbertsmithite}\label{appB}

In Table~\ref{tab:JherbertsmithiteU} we provide the  results of total energy calculations with GGA+U functional
using different values of $U$ for {\zn}.

\begin{table}[htb]
  \caption{ Exchange coupling constants for kapellasite determined from total energies of 
    five different spin configurations in a $2\times2\times1$ supercell.
  }\label{tab:Jkapellasite221}
\begin{ruledtabular}
\begin{tabular}{cD{.}{.}{1.5}cD{.}{.}{1.1}D{.}{.}{1.1}D{.}{.}{1.1}}
name& \multicolumn{1}{c}{$d_{Cu-Cu}$}&type&\multicolumn{1}{c}{$J_i$~(K)}&\multicolumn{1}{c}{$J_i$~(K)}&\multicolumn{1}{c}{$J_i$~(K)}\\
&&&\multicolumn{1}{c}{$U\!=\!6\,$eV}&\multicolumn{1}{c}{$U\!=\!7\,$eV}&\multicolumn{1}{c}{$U\!=\!8\,$eV}\\\hline
$J_1$&3.15& {\ka} nn& -14.2 &-13.4&-12.6\\
$J_2$&5.45596&{\ka} 2nd nn&  -0.7&-0.7&-0.6\\
$J_4$&6.3 & {\ka} 3rd nn& -0.3&-0.3&-0.3\\
$J_d$&6.3 & {\ka} 3rd nn& 24.0&19.8&16.3
\end{tabular}
\end{ruledtabular}
\end{table}

\begin{table}[htb]
  \caption{ Exchange coupling constants for kapellasite determined from total energies of 
    eight different spin configurations in a $2\times1\times2$ supercell. Note that here, 
    {\ka} 3rd nearest neigbours are symmetry equivalent so that $J_4'=\nicefrac{2}{3}J_4+\nicefrac{1}{3}J_d$.
  }\label{tab:Jkapellasite212}
\begin{ruledtabular}
\begin{tabular}{cD{.}{.}{1.5}cD{.}{.}{1.1}D{.}{.}{1.1}D{.}{.}{1.1}}
name& \multicolumn{1}{c}{$d_{Cu-Cu}$}&type&\multicolumn{1}{c}{$J_i$~(K)}&\multicolumn{1}{c}{$J_i$~(K)}&\multicolumn{1}{c}{$J_i$~(K)}\\
&&&\multicolumn{1}{c}{$U\!=\!6\,$eV}&\multicolumn{1}{c}{$U\!=\!7\,$eV}&\multicolumn{1}{c}{$U\!=\!8\,$eV}\\\hline
\multicolumn{6}{c}{{\bf {\ka} layer couplings}}\\\hline
$J_1$&3.15& {\ka} nn& -11.8 &-11.3&-10.9\\
$J_2$&5.45596&{\ka} 2nd nn&  -1.2&-1.0&-0.9\\
$J_4'$&6.3 & {\ka} 3rd nn& 10.8&8.8&7.3\\\hline
\multicolumn{6}{c}{{\bf interlayer couplings}}\\\hline
$J_3$&5.733& interlayer 1st nn&  -0.4&-0.3&-0.2\\
$J_5$&6.54139& interlayer 2nd nn&  0.04&0.04&0.04\\
$J_6$&7.91421& interlayer 3rd nn&  0.04&0.04&0.03\\
$J_8$&8.51806& interlayer 4th nn&  0.01&<0.01&<0.01\\
\end{tabular}
\end{ruledtabular}
\end{table}

\begin{table}[htb]
  \caption{ 
    Exchange coupling constants for {\zn} (herbertsmithite) determined 
    from total energies of nine different spin configurations. Energies 
    were calculated with GGA+U functionals at $J=1$~eV with different values of $U$ and 
    with atomic limit double counting correction.}\label{tab:JherbertsmithiteU}
\begin{ruledtabular}
\begin{tabular}{cD{.}{.}{1.5}cD{.}{.}{1.1}D{.}{.}{1.1}D{.}{.}{1.1}}
name& \multicolumn{1}{c}{$d_{Cu-Cu}$}&type&\multicolumn{1}{c}{$J_i$~(K)}&\multicolumn{1}{c}{$J_i$~(K)}&\multicolumn{1}{c}{$J_i$~(K)}\\
&&&\multicolumn{1}{c}{$U\!=\!6\,$eV}&\multicolumn{1}{c}{$U\!=\!7\,$eV}&\multicolumn{1}{c}{$U\!=\!8\,$eV}\\\hline
\multicolumn{6}{c}{{\bf {\ka} layer couplings}}\\\hline
$J_1$&3.4171& {\ka} nn&182.4  &155.4&131.8\\
$J_3$&5.91859&{\ka} 2nd nn&  3.4 &2.9&2.3\\
$J_5$&6.8342 & {\ka} 3rd nn& -0.4 &-0.5&-0.4\\\hline
\multicolumn{6}{c}{{\bf interlayer couplings}}\\\hline
$J_2$&5.07638& interlayer 1st nn&  5.3 &4.5&3.7\\
$J_4$&6.11933& interlayer 2nd nn& -1.5 &-1.1&-0.8\\
$J_6$&7.00876& interlayer 3rd nn&  -6.4 &-5.4&-4.4\\
$J_7$&8.51328& interlayer 4th nn&  3.0  &2.5&2.1\\
$J_9$&9.17347& interlayer 6th nn&  2.5  &2.1&1.7\\
\end{tabular}
\end{ruledtabular}
\end{table}


\begin{thebibliography}{99}

\bibitem{Balents2010} L. Balents, Nature {\bf 464}, 199 (2010).

\bibitem{Shores2005} M. P. Shores, E. A. Nytko, B. M. Bartlett and
  D. G. Nocera, J. Am. Chem. Soc. {\bf 127}, 13462 (2005).

\bibitem{Lee2008} P. A. Lee, Science {\bf 321}, 1306 (2008).

\bibitem{Mendels2010} P. Mendels and F. Bert, J. Phys. Soc. Jpn. {\bf
    79}, 011001 (2010).

\bibitem{Mendels2011} P. Mendels and F. Bert, J. Phys.: Conf. Series
  {\bf 320}, 012004 (2011).

\bibitem{Helton2007} J. S. Helton, K. Matan, M. P. Shores,
  E. A. Nytko, B. M. Bartlett, Y. Yoshida, Y. Takano, A. Suslov,
  Y. Qiu, J.-H. Chung, D. G. Nocera, and Y. S. Lee, Phys. Rev. Lett.
  {\bf 98}, 107204 (2007).

\bibitem{Mendels2007} P. Mendels, F. Bert, M. A. de Vries, A. Olariu,
  A. Harrison, F. Duc, J. C. Trombe, J. S. Lord, A. Amato, and
  C. Baines, Phys. Rev. Lett.  {\bf 98}, 077204 (2007).
 
\bibitem{Vries2009} M. A. de Vries, J. R. Stewart, P. P. Deen,
 J. O. Piatek, G. J. Nilsen, H. M. Ronnow, and A. Harrison,
Phys. Rev. Lett. {\bf 103}, 237201 (2009).
 
\bibitem{Han2012} T.-H. Han, J. S. Helton, S. Chu, D. G. Nocera,
  J. A. Rodriguez-Rivera, C. Broholm, Y. S. Lee, Nature {\bf 492}, 406
  (2012).

\bibitem{Olariu2008} A. Olariu, P. Mendels, F. Bert, F. Duc,
  J. C. Trombe, M. A. de Vries, and A. Harrison, Phys. Rev. Lett.
  {\bf 100}, 087202 (2008).

\bibitem{Freedman2010} D. E. Freedman, T. H. Han, A. Prodi, Peter
  M{\"u}ller, Q.-Z. Huang, Y.-S. Chen, S. M. Webb, Y. S. Lee,
  T. M. McQueen, and D. G. Nocera, J. Am. Chem. Soc. {\bf 132}, 16185
  (2010).

\bibitem{Han2012b} T. Han, S. Chu and Y. S. Lee, Phys. Rev. Lett.
  {\bf 108}, 157202 (2012).

\bibitem{Rigol2007} M. Rigol and R. R. P. Singh, Phys. Rev. Lett.
  {\bf 98}, 207204 (2007).

\bibitem{Messio2010} L. Messio, O. C\'epas, and C. Lhuillier,
Phys. Rev. B {\bf 81}, 064428 (2010).


\bibitem{Zorko2008} A. Zorko, S. Nellutla, J. van Tol, L. C. Brunel,
  F. Bert, F. Duc, J.-C. Trombe, M. A. de Vries, A. Harrison, and
  P. Mendels, Phys. Rev. Lett. {\bf 101}, 026405 (2008).

\bibitem{Shawish2010}
S. El Shawish, O. Cepas, and S. Miyashita,
Phys. Rev. B {\bf 81}, 224421 (2010).


\bibitem{Koepernik1999} K. Koepernik and H. Eschrig, Phys. Rev. B
 {\bf 59}, 1743 (1999); {\tt http://www.FPLO.de}.

\bibitem{Perdew1996} J. P. Perdew, K. Burke and M. Ernzerhof,
  Phys. Rev. Lett. {\bf 77} 3865 (1996).

\bibitem{Jeschke2011} H. O. Jeschke, I. Opahle, H. Kandpal,
 R. Valent\'{i}, H. Das, T. Saha-Dasgupta, O. Janson, H. Rosner,
 A. Br\"{u}hl, B. Wolf, M. Lang, J. Richter, S. Hu, X. Wang,
 R. Peters, T. Pruschke, and A. Honecker, Phys. Rev. Lett. {\bf
 106}, 217201 (2011).

\bibitem{Janson2008} O. Janson, J. Richter and H. Rosner,
  Phys. Rev. Lett. {\bf 101}, 106403 (2008).

\bibitem{Fak2012}
  B. F{\aa}k, E. Kermarrec, L. Messio, B. Bernu, C. Lhuillier,
  F. Bert, P. Mendels, B. Koteswararao, F. Bouquet, J. Ollivier,
  A. D. Hillier, A. Amato, R. H. Colman, and A. S. Wills,
  Phys. Rev. Lett.  {\bf 109}, 037208 (2012).

\bibitem{Krause2006} W. Krause, H.-J. Bernhardt, R. S. W. Braithwaite,
  U. Kolitsch, and R. Pritchard, Mineralog. Mag. {\bf 70}, 329 (2006).

\bibitem{hydrogen} For relaxation, we use the GGA functional,
  $10\times10\times10$ $k$ mesh and optimize only the H position. We
  obtain the Wyckoff position
  $(-0.1993636755,0.1993636755,-0.1774445876)$. The H-O distance is
  $d=0.993$~{\AA}.


\bibitem{Foyevtsova2011}
K. Foyevtsova, I. Opahle, Y.-Z. Zhang, H. O. Jeschke, Roser Valent\'\i,
Phys. Rev. B {\bf 83}, 125126 (2011).

\bibitem{Bernu2013} B. Bernu, C. Lhuillier, E. Kermarrec, F. Bert, and
  P. Mendels, Phys. Rev. B {\bf 87}, 155107 (2013).

\bibitem{Tay2012} T. Tay, O. I. Motrunich, Phys. Rev. B {\bf 84},
 020404(R) (2011).

\bibitem{Iqbal2012} Y. Iqbal, F. Becca, and D. Poilblanc, New
  J. Phys. {\bf 14}, 115031 (2012).

\bibitem{Messio2012} L. Messio, B. Bernu, C. Lhuillier,
  Phys. Rev. Lett. {\bf 108}, 207204 (2012).

\bibitem{Schmalfuss2004} D. Schmalfu\ss, J. Richter, and D. Ihle,
Phys. Rev. B {\bf 70}, 184412 (2004).

\bibitem{Schmidt2011} H.-J. Schmidt, A. Lohmann, and J. Richter,
  Phys. Rev. B {\bf 84}, 104443 (2011); high temperature series
  expansion code from {\tt http://www.uni-magdeburg.de/jschulen/HTE/}.

\bibitem{Misguich2007} G. Misguich and P. Sindzingre, Eur. Phys. J. B
  {\bf 59}, 305 (2007).

\bibitem{Singh2009} R. R. P. Singh and M. Rigol, J. Phys: Conf. Series
  {\bf 145}, 012003 (2009).

\bibitem{Suttner2013} R. Suttner, C. Platt, J. Reuther, and
  R. Thomale, Phys. Rev. B {\bf 89}, 020408(R) (2014).


\bibitem{Zhang2008}
Y.-Z. Zhang, H. O. Jeschke, R. Valent\'\i,
Phys. Rev. B {\bf 78}, 205104 (2008).

\bibitem{Momma2011} K. Momma and F. Izumi,  J. Appl. Crystallogr. {\bf 44}, 1272 (2011).

\end{thebibliography}
\end{document}